\documentclass[10pt]{article}

\usepackage{fullpage}
\usepackage{algorithm}
\usepackage{algorithmic}
\usepackage{latexsym}
\usepackage{amsfonts,amsmath}

\def\01{\{0,1\}}
\newcommand{\ket}[1]{\vert #1 \rangle}

\newcommand{\set}[1]{\{ #1 \}}

\newtheorem{Theorem}{Theorem}
\newtheorem{Lemma}{Lemma}
\newtheorem{Definition}{Definition}
\newtheorem{Corollary}{Corollary}

\newenvironment{Proof}{\noindent
$\mathbf{Proof.}$}{\hspace{\stretch{1}}$\Box$\\ \smallskip}

\begin{document}

\author{Mart~de~Graaf \and Ronald~de~Wolf}
\date{}
\title{On Quantum Versions of the Yao Principle\thanks{CWI, INS4, P.O.~Box 94079,
1090 GB Amsterdam, The Netherlands. Email: $\mathtt{\{mgdgraaf,rdewolf\}@cwi.nl}$.
Partially supported by the EU fifth framework project QAIP, IST--1999--11234.
Mart de Graaf is also supported by grant 612.055.001 from the Netherlands
Organization for Scientific Research (NWO). Ronald de Wolf is also supported
by NWO TALENT grant S 62-565.}}

\maketitle


\begin{abstract}
The classical Yao principle states that the complexity $R_\epsilon(f)$ of
an optimal \emph{randomized} algorithm for a function $f$ with
success probability $1-\epsilon$ equals the complexity $\max_\mu D_\epsilon^\mu(f)$ 
of an optimal \emph{deterministic} algorithm for $f$ that is correct
on a fraction $1-\epsilon$ of the inputs, weighed according to
the hardest distribution $\mu$ over the inputs. In this paper we investigate
to what extent such a principle holds for quantum algorithms.
We propose two natural candidate quantum Yao principles, a ``weak'' and a
``strong'' one. For both principles, we prove that the quantum bounded-error 
complexity is a lower bound on the quantum analogues of $\max_\mu D_\epsilon^\mu(f)$.
We then prove that equality cannot be obtained for the ``strong'' version,
by exhibiting an exponential gap. On the other hand, as a positive result
we prove that the ``weak'' version holds up to a constant factor for the 
query complexity of all symmetric Boolean functions.\\[2mm] 
\end{abstract}


\section{Introduction}

\subsection{Motivation}
In classical computing, the \emph{Yao principle}~\cite{yao:unified}
gives an equivalence between two kinds of randomness in algorithms:
randomness inside the algorithm itself, and randomness on the inputs.
Let us fix some model of computation for computing a Boolean function $f$,
like query complexity, communication complexity, etc.
Let $R_\epsilon(f)$ be the minimal complexity among all
\emph{randomized} algorithms that compute $f(x)$ with success
probability at least $1-\epsilon$, for all inputs $x$.
Let $D_\epsilon^\mu(f)$ be the minimal complexity among all
\emph{deterministic} algorithms that compute $f$ correctly on
a fraction of at least $1-\epsilon$ of all inputs, weighed
according to a distribution $\mu$ on the inputs.
The Yao principle now states that these complexities
are equal if we look at the ``hardest'' input distribution $\mu$:
$$
R_\epsilon(f) = \max_\mu D_\epsilon^\mu(f).
$$
This is a special case of Von Neumann's minimax theorem in
game theory~\cite{neumann47theory,owen:gametheory}.

Since its introduction, the Yao principle has been an extremely
useful tool in computational complexity analysis.
In particular, it allows us to 
derive lower bounds on randomized algorithms from lower bounds 
on deterministic algorithms: choose some ``hard'' input distribution
$\mu$, prove a lower bound on deterministic algorithms that compute
$f$ correctly for ``most'' inputs, weighted according to $\mu$, and 
then use  $R_\epsilon(f)\geq D_\epsilon^\mu(f)$ to get a lower bound on
$R_\epsilon(f)$.
This method is used very often, because it is usually much easier to analyze 
deterministic algorithms than to analyze randomized ones.

In recent years \emph{quantum computation} received a lot of 
attention. Here quantum mechanical principles are 
employed to realize more efficient computation than is  
possible with a classical computer.  Famous examples are Shor's 
polynomial-time factoring algorithm~\cite{shor:factoring} and 
Grover's search algorithm~\cite{grover:search}. 
However, the field is still young and open questions are abundant. In
particular, there has been a search for good techniques to provide
lower bounds on quantum algorithms. Most of these lower bounds
are in the \emph{query model}, where the complexity of an algorithm 
is measured by the number of queries it needs 
in order to compute some function
(we will provide formal definitions of this and other concepts 
in the next section). Two general methods in this direction
are the \emph{polynomial method} introduced by Beals, Buhrman, Cleve, 
Mosca, and de Wolf \cite{bbcmw:polynomials} and the method of 
\emph{quantum adversaries} of Ambainis \cite{ambainis:lowerbounds}. In this
paper we investigate the possibility of a third method, a \emph{quantum
Yao principle}. It is our hope that such a principle will prove itself
useful as a link between techniques for lower bounds on exact and 
bounded-error quantum algorithms.

The first difficulty one runs into when investigating a quantum
version of the Yao principle, is the question what the proper 
quantum counterparts of $R_\epsilon(f)$ and $D_\epsilon^\mu(f)$ are.
Let us fix the error probability at $\epsilon=\frac{1}{3}$ here
(any other value in $(0,\frac{1}{2})$ would do as well).
The quantum analogue of $R_{1/3}(f)$ is straightforward:
let $Q_2(f)$ denote the minimal complexity among all
\emph{quantum} algorithms that compute $f(x)$ with probability 
at least $\frac{2}{3}$, for all inputs $x$.
However, the inherently ``random'' nature of quantum algorithms prohibits 
a straightforward definition of ``deterministic'' quantum algorithms 
in analogy of deterministic classical algorithms. 
We therefore propose two different definitions,
a weak and a strong one. In the following, let $f:D\rightarrow\01$ 
be some function that we want to compute, with $D\subseteq\01^N$.
If $D=\set{0,1}^N$ then $f$ is a \emph{total} function, 
otherwise $f$ is a \emph{promise} function.
Let $A$ be a quantum algorithm, $P_A(x)$ the acceptance probability of $A$ 
on input $x$ (the probability of outputting 1 on input $x$, and 
$\mu:D\rightarrow[0,1]$ a probability distribution over the inputs. 

\begin{Definition}
$A$ is \emph{weakly $\frac{2}{3}$-exact} for $f$ with respect to $\mu$ 
iff $\mu(\set{x \mid P_A(x)=f(x)}) \geq \frac{2}{3}$.
\end{Definition}

\begin{Definition}
$A$ is \emph{strongly $\frac{2}{3}$-exact} for $f$ with respect to $\mu$ 
iff $A$ is weakly $\frac{2}{3}$-exact for $f$ with respect to $\mu$ 
and $P_A(x) \in \set{0,1}$ for all inputs $x \in \set{0,1}^N$.
\end{Definition}
Informally, in the second definition we require the algorithm
to output the same output on the same input, even on inputs $x\in D$ 
where the algorithm fails and even on $x\in\set{0,1}^N\backslash D$
(similar to a classical deterministic algorithm). 
In the first definition, we only require this 
``input-determines-output'' behavior to occur for a $\mu$-fraction
of at least $\frac{2}{3}$ of the inputs where the algorithm 
gives the correct output $f(x)$.
Note that a strongly $\frac{2}{3}$-exact algorithm for $f$ with 
respect to $\mu$ actually computes some total function 
$g:\set{0,1}^N \rightarrow\01$ with success probability 1, 
namely the function $g(x)=P_A(x)$. This $g$ will agree with $f$
on at least $\frac{2}{3}$ of the inputs.

These two definitions lead to a weak and a strong quantum counterpart
to the classical distributional complexity $D^\mu_{1/3}(f)$:
let $Q_{WE}^\mu(f)$ and $Q_{SE}^\mu(f)$ denote the minimal complexity 
among all weakly and strongly $\frac{2}{3}$-exact algorithms for $f$ 
with respect to $\mu$, respectively.
We can now state two potential quantum versions of the Yao principle:
\begin{itemize}
\item Strong quantum Yao principle: 
$\displaystyle Q_2(f) \stackrel{?}{=} \max_\mu Q_{SE}^\mu(f)$ 
\item Weak quantum Yao principle:
$\displaystyle Q_2(f) \stackrel{?}{=} \max_\mu Q_{WE}^\mu(f)$ 
\end{itemize}
In this paper we investigate to what extent these two quantum
Yao principles hold.

\subsection{Results}
Our results are threefold. Firstly, we prove that both of these principles
hold in the `$\leq$'-direction, for all $f$:
\begin{itemize}
\item $\displaystyle Q_2(f) \leq \max_\mu Q_{SE}^\mu(f)$
\item $\displaystyle Q_2(f) \leq \max_\mu Q_{WE}^\mu(f)$
\end{itemize}
Clearly, the second inequality implies the first,
since $Q_{WE}^\mu(f)\leq Q_{SE}^\mu(f)$ for all $f$ and $\mu$.
The proof is similar to the classical game-theoretic proof,
with a bit more technical complication.
We emphasize that this result is perfectly general,
and applies to all computational models to which the classical
Yao principle applies.

In order to investigate to what extent the `$\geq$'-directions 
of these two quantum Yao principles hold, we instantiate our complexity 
measures to the query complexity setting. Our second result
is an exponential gap between $Q_2(f)$ and $Q_{SE}^\mu(f)$
for the query complexity of Simon's problem~\cite{simon:power}:
\begin{itemize}
\item There exist $f$ and  $\mu$ such that 
$Q_2(f)$ is exponentially smaller than $Q_{SE}^\mu(f)$.
\end{itemize}
This shows that the strong quantum Yao principle is false.
Thirdly, we prove that the weak quantum Yao principle holds up to
a constant factor for the query complexity of all \emph{symmetric} functions:
\begin{itemize}
\item $\displaystyle Q_2(f)=\Theta\left(\max_\mu Q_{WE}^\mu(f)\right)$ for all symmetric $f$
\end{itemize}
For this result we first construct a quantum algorithm that
can determine the $N$-bit input $x$ \emph{with certainty} in $O(\sqrt{kN})$
queries if $k$ is a known upper bound on the Hamming weight of $x$.
We then use that algorithm to construct, for every symmetric function $f$ 
and distribution $\mu$, a quantum algorithm that computes $f(x)$
with certainty for ``most'' inputs $x$. In addition to this result
for symmetric functions, we also show that for a particular \emph{monotone}
non-symmetric function $f$, the $\max_\mu Q_{WE}^\mu(f)$ complexity lies 
in between the best known bounds for $Q_2(f)$.


\section{Preliminaries}

In this section we formalize the notion of query complexity,
define several complexity measures, state Von Neumann's 
minimax theorem and derive the classical Yao principle from it.

\subsection{Query Complexity}
We assume familiarity with classical computation theory and briefly 
sketch the basics of quantum computation; an extensive introduction may
be found in the book by Nielsen and Chuang \cite{nielsen&chuang:qc}.
Quantum algorithms operate on \emph{qubits} as opposed to bits
in classical computers. The state of an $m$-qubit quantum system
can be written as
$$
\ket{\phi} = \sum_{i \in \set{0,1}^m} \alpha_i \ket{i},
$$
where $\ket{i}$ denotes the basis state $i$, which is a classical
$m$-bit string. The $\alpha_i$'s are complex numbers known as the 
\emph{amplitudes} of the basis states $\ket{i}$ and we require
$\sum_{i\in \set{0,1}^m} |\alpha_i|^2 = 1$. Mathematically, the
state of a system is thus described by a $2^m$-dimensional complex unit vector.
If we measure the value of $\ket{\phi}$, then we will see the basis 
state $\ket{i}$ with probability $|\alpha_i|^2$, after which 
the system collapses to $\ket{i}$. Operations which are not 
measurements on a system of qubits correspond to 
\emph{unitary transformations} on the vector of amplitudes. 

In the \emph{query model} of computation, the goal is to compute
some function $f:D \to \set{0,1}$ on an input $x\in D\subseteq\set{0,1}^N$, 
using as few accesses (``queries'') to the $N$ input bits as possible.
In quantum algorithms, it is by now standard to formalize a query as
an application of a unitary transformation $O$ that acts as follows:
$$
O\ket{i,b,z} = \ket{i,b \oplus x_i,z}.
$$
Here $i \in \set{1,\ldots,N}$, $b\in\set{0,1}$, 
$\oplus$ denotes the exclusive-or 
function, and $z$ denotes the workspace of the algorithm, 
which is not affected by $O$.
A $T$-query quantum algorithm $A$ then has the form 
$$
A=U_TOU_{T-1}O\cdots U_1OU_0,
$$
with each $U_i$ a fixed unitary transformation independent of the
input $x$. $A$ is assumed to start in the all-zero state $\ket{0\ldots0}$,
and its output (0 or 1) is obtained by measuring the rightmost bit of its
final state $A\ket{0\ldots0}$.
The \emph{acceptance probability} $P_A(x)$ of a quantum algorithm
$A$ is defined as the probability of getting output $1$ on input $x$.
Its \emph{success probability} $S_A(x)$ is the probability of getting 
the correct output $f(x)$ on input $x$.

A quantum algorithm $A$ computes a function $f:D\rightarrow\set{0,1}$ 
\emph{exactly} if $S_A(x)=1$ for all inputs $x\in D$. 
Algorithm $A$ computes $f$ with \emph{bounded-error} if $S_A(x) \geq \frac{2}{3}$ 
for all $x\in D$. We use $Q_E(f)$ and $Q_2(f)$ to denote the minimal number
of queries required by exact and bounded-error quantum algorithms for 
$f$, respectively. These complexities are the quantum versions 
of the classical deterministic and
bounded-error decision tree complexities $D(f)$ and $R_2(f)$, respectively.
For completeness, we repeat our two alternative quantum
versions of the classical distributional complexity $D^\mu(f)$ from the introduction.
Let $\mu$ be a probability distribution on the set of all possible inputs.  
An algorithm $A$ is \emph{weakly $\frac{2}{3}$-exact for $f$ with respect 
to $\mu$} if $\mu(\set{x \mid P_A(x) = f(x)}) \geq \frac{2}{3}$, and $A$ 
is \emph{strongly $\frac{2}{3}$-exact for $f$ with respect to $\mu$} if
$A$ is weakly $\frac{2}{3}$-exact for $f$ with respect to $\mu$ and
$P_A(x) \in \set{0,1}$ for all $x\in\set{0,1}^N$. By $Q_{SE}^\mu(f)$ and
$Q_{WE}^\mu(f)$ we denote the minimal number of queries needed by
strongly and weakly $\frac{2}{3}$-exact quantum algorithms
for $f$ with respect to $\mu$, respectively. 
Note that $Q_{WE}^\mu(f)\leq Q_{SE}^\mu(f)$ for all $f$ and $\mu$, 
hence in particular $\max_\mu Q_{WE}^\mu(f)\leq\max_\mu Q_{SE}^\mu(f)$.

One of the first quantum algorithms operating in the query model
is Grover's search algorithm \cite{grover:search,bhmt:countingj}.
If $t=|x|>0$ then the algorithm uses $\frac{\pi}{4}\sqrt{N/t}$ 
queries and with high probability outputs an $i$ such that $x_i=1$. 
Here we use $|x|$ to denote the Hamming weight (number of 1's) in $x$,
and $x_i$ to denote the $i$th bit of $x$.
If $|x|=0$ then the algorithm outputs `no solutions'.
Brassard, H\o yer, Mosca, and Tapp~\cite{bhmt:countingj} give 
an exact version of Grover's algorithm that can accomplish the same 
task with probability 1 if $t$ (the number of 1's in the input) is known.

For $\emph{total}$ functions $f:\set{0,1}^N \to \set{0,1}$, Beals,
Buhrman, Cleve, Mosca, and de Wolf \cite{bbcmw:polynomials}
proved that classical deterministic query complexity $D(f)$ is polynomially
related to the exact and bounded-error quantum complexities:
$D(f) = O(Q_E(f)^4)$ and $D(f) = O(Q_2(f)^6)$. 

A function $f:\set{0,1}^N \to \set{0,1}$ is \emph{symmetric}
if its value $f(x)$ depends only on $|x|$. For such $f$, 
define $f_k = f(x)$ where $|x|=k$. In \cite{bbcmw:polynomials} it is 
proven that $Q_2(f) = \Theta(\sqrt{N(N-\Gamma(f))})$, where $\Gamma(f) = 
\min \set{ |2k-N-1| \mid f_k \ne f_{k+1} \mathrm{~and~} 0 \leq k \leq N-1}$.
Informally, the quantity $\Gamma(f)$ measures the length of the interval
around Hamming weight $\frac{N}{2}$ where $f$ is constant. 
A symmetric function $f$ is a \emph{threshold}
function if there is a $0 < t \leq N$, such that $f(x)=1$ iff
$|x| \geq t$. Note that for $t\leq N/2$ we have 
$Q_2(f)=\Theta(\sqrt{tN})$ as a direct consequence
of the bound for symmetric functions. A function $f:\01^n \to \01$ is 
\emph{monotone} if $(\forall i\ x_i \leq y_i) \Rightarrow f(x) \leq f(y)$.

\subsection{The Classical Yao Principle}

Consider the following game-theoretic setting: 
player~1 has a choice between some $m$ ``pure'' strategies 
and player~2 has a choice between $n$ ``pure'' strategies.
If player~1 plays $i$ and player~2 plays $j$,
then player~1 receives ``payoff'' $P_{ij}$. 
Player~1 wants to maximize the payoff, player~2 wants to minimize.
Viewing $P$ as an $m\times n$ matrix, and using $e_i$ and $e_j$
to denote the appropriate unit column vectors with a 1 in place $i$, 
respectively $j$, the payoff corresponds to the matrix product $e_i^TPe_j$.
However, the players may also use ``mixed'' strategies (probability 
distributions over ``pure'' strategies) to further their goals.
Mixed strategies of players~1 and 2 correspond to $m$- and $n$-dimensional
column vectors $\rho$ and $\mu$, respectively, of non-negative reals that sum to 1.
Now the \emph{expected} payoff is $\rho^TP\mu$.
Note that if player~1 can choose his strategy $\rho$ knowing player~2's
strategy $\mu$, 
then he would choose $\rho$ to maximize the payoff $\rho^TP\mu$; 
in this situation player~2 would do best to choose $\mu$ to minimize
$\max_{\rho}\rho^TP\mu$, 
giving expected payoff $\min_\mu\max_\rho \rho^TP\mu$.
Conversely, if player~2 could choose his strategy knowing player~1's strategy,
then the expected payoff would be $\max_\rho\min_\mu\rho^TP\mu$.
Von~Neumann's famous minimax theorem~\cite{neumann47theory,owen:gametheory} 
tells us that these two quantities are in fact equal:
$$
\min_\mu \max_\rho \rho^TP\mu = \max_\rho \min_\mu \rho^TP\mu.
$$
It is not hard to see that without loss of generality the ``inner'' choices 
can be assumed to be pure strategies, so as an easy consequence we also have
$$
\min_\mu \max_i e_i^TP\mu = \max_\rho \min_j \rho^TPe_j.
$$
Yao~\cite{yao:unified} was the first to interpret this result in
computational terms. We will sketch the computational interpretation below.
Fix some classical model of computation for which the set of deterministic 
algorithms of complexity $\leq c$ is finite, for every $c$.
Examples of such models are query complexity, communication complexity,
 etc. Player~1 chooses an algorithm to compute 
$f:D\rightarrow\set{0,1}$ and player~2 chooses an input $x$ that 
is hard for player~1. 
The pure strategies for player~1 are all \emph{deterministic} classical
algorithms of complexity $\leq c$ and hence his mixed strategies are all 
\emph{randomized} classical algorithms of complexity $\leq c$.
The pure strategies for player~2 are the inputs in $D$ and his mixed
strategies are all probability distributions $\mu$ over $D$.
We define the payoff matrix such that $P_{ix}=1$ if algorithm $i$ computes $f$ 
correctly on input $x$, and $P_{ix}=0$ otherwise. 
In this setting, the minimax theorem states
$$
\min_\mu \max_i e_i^TP\mu = \max_\rho \min_x \rho^TPe_x.
$$
Let us interpret both sides of this equation.
On the left, the quantity $e_i^TP\mu$ is the fraction of inputs 
on which deterministic algorithm $i$ is correct, weighed according to $\mu$,
and $\max_i e_i^TP\mu$ denotes this fraction for the optimal deterministic
algorithm of complexity $\leq c$. Thus the left-hand-side of the equation
gives this optimal correct fraction for the hardest distribution $\mu$
achievable by deterministic complexity-$c$ algorithms.
On the other hand, $\rho^TPe_x$ is the success probability on input $x$
achieved by the randomized algorithm given by probability distribution $\rho$ 
over deterministic algorithms, and $\min_x \rho^TPe_x$ is its success
probability on the hardest input. Thus the right-hand-side gives the
highest worst-case success probability achievable by 
randomized complexity-$c$ algorithms. Since these two quantities are equal
for all $c$, we obtain the classical Yao principle:
$$
R_\epsilon(f) = \max_\mu D_\epsilon^\mu(f).
$$


\section{Proof of One Half of the Quantum Yao Principle}

As a first result we prove that $Q_2(f) \leq \max_{\mu}Q_{WE}^{\mu}(f)$. 
The proof is similar to the derivation of the classical Yao 
principle above, but the details are a bit more messy. 

\begin{Theorem}
\label{thm:Q2lowerbound}
For all $f:D \to \set{0,1}$, with $D$ finite,
$\displaystyle Q_2(f) \leq \max_{\mu}Q_{WE}^{\mu}(f)$.
\end{Theorem}

\begin{Proof}
Consider the (infinite) set of all quantum algorithms of complexity 
$\leq \max_{\mu}Q_{WE}^\mu(f)$.
Let $i$ be any algorithm from this set, and $x \in D$ an input. 
Consider the quantity $\lfloor S_i(x) \rfloor$, which is 1 if algorithm $i$
computes $f(x)$ with success probability 1, and which is 0 otherwise.
Call algorithms $i$ and $j$ \emph{similar} if 
$\lfloor S_i(x) \rfloor=\lfloor S_j(x) \rfloor$ for all $x\in D$.
In this way, similarity is an equivalence relation on the set of 
all quantum algorithms of complexity $\leq \max_{\mu}Q_{WE}^\mu(f)$. 
Note that this relation has at most $2^{|D|}$ equivalence classes.
From each equivalence class, we choose as a representative an algorithm 
from that class with the least complexity. 

Now consider the game in which player 1 wants to compute $f$, and as 
pure strategies he has 
available the (finite) set of representatives of the equivalence classes. 
Player 2 is an adversary that tries to make life as hard as 
possible for player 1 by choosing hard inputs $x\in D$ to $f$. 
Let $S$ be the matrix of success probabilities ($S_{ix}=S_i(x)$).
Define the payoff matrix as $P_{ix} = \lfloor S_{ix} \rfloor$.
Now consider the quantity $\max_i e_i^TP\mu$. 
This represents the $\mu$-fraction of inputs on which the best 
weakly $\frac{2}{3}$-exact quantum algorithm for $f$ with respect to
that $\mu$ is correct. By construction, this quantity is at least 
$\frac{2}{3}$ for all $\mu$.
Using the minimax theorem, we now obtain:
$$
\frac{2}{3} \leq \min_\mu \max_i e_i^T P\mu = 
\max_\rho \min_x \rho^T P e_x \leq \max_\rho \min_x \rho^T S e_x.
$$
Here the last term can be interpreted as the success probability
of a quantum algorithm formed by a probability distribution $\rho$ over 
the set of representatives of the equivalence classes. By the above
inequality, this algorithm has success probability 
$\geq\frac{2}{3}$ for all inputs $x\in D$.
Since it is a probability distribution over algorithms of complexity
$\leq \max_{\mu}Q_{WE}^\mu(f)$, its complexity is at most 
$\max_{\mu}Q_{WE}^\mu(f)$. Hence \hbox{$Q_2(f) \leq \max_\mu Q_{WE}^\mu(f)$.}
\end{Proof}

\begin{Corollary}
\label{cor:Q2lowerbound}
For all $f:D \to \set{0,1}$, with $D$ finite,
$\displaystyle Q_2(f) \leq \max_{\mu}Q_{SE}^{\mu}(f)$.
\end{Corollary}

Note that although we restrict our attention to the query model of
computation, the proofs of Theorem \ref{thm:Q2lowerbound} and Corollary 
\ref{cor:Q2lowerbound} also work for the other models of complexity 
where the classical Yao principle applies.


\section{A Counterexample for the Strong Quantum Yao Principle}

In this section we prove that the strong quantum Yao principle does not hold. 
There exists a problem $f$ such that for a suitable distribution 
$\mu$, $Q_2(f)$ is exponentially smaller than $Q_{SE}^\mu(f)$.
This exponential gap follows from a known result about 
the classical and quantum complexity of 
\emph{Simon's problem} \cite{simon:power}, and the fact that classical 
deterministic and quantum exact complexity are polynomially related for 
total problems \cite[Theorem 5.4]{bbcmw:polynomials}. 

\begin{Theorem}
\label{thm:exponentialGap}
There exist a problem $f$ and a distribution $\mu$ such that $Q_2(f)
=O(n^2)$ and $Q_{SE}^{\mu}(f)=\Omega(2^{\frac{n}{8}})$.
\end{Theorem}

\begin{Proof}
Consider Simon's problem: given a function
$\phi:\{0,1\}^n \to \{0,1\}^n$ with the promise that there is
an $s \in \{0,1\}^n$ such that $\phi(a)=\phi(b)$ iff $a \oplus b =s$,
decide whether $s=0$ or not. This function $\phi$ is given as an input
$x$ of $N=n2^n$ bits, using $n$ 1-bit entries for each function
value $\phi(\cdot)$. The input bits can be queried in the usual way.
Using Simon's bounded-error quantum algorithm, this problem can 
be solved in $O(n^2)$ queries, and hence $Q_2(Simon) =
O(n^2)$. Now define a distribution $\mu$ which uniformly places half the 
total weight on inputs with $s=0$ and half the total weight on inputs 
with $s \ne 0$:
$$
\mu(x) = \left\{ \begin{array}{ll}
\frac{1}{2(2^n)!} & \textrm{if $s = 0$} \\
\frac{1}{2(2^n-1){2^n \choose 2^{n-1}}(2^{n-1})!} &
\textrm{if $s \ne 0$}\\
0 & \textrm{else.}
\end{array} \right.
$$
Simon proved that under this distribution, any classical algorithm that is correct on
a fraction $\geq \frac{2}{3}$ requires $\Omega(\sqrt{2^n})$ queries.
Now take any strongly $\frac{2}{3}$-exact quantum algorithm $A$ that solves 
this problem and makes $T$ queries, then $A$ 
computes some \emph{total} function $g$. Since $D(g) = O(Q_E(g)^4)$, 
this implies that there exists a deterministic classical algorithm
that computes $g$ using $O(T^4)$ queries. But this classical algorithm
is then exact on a $\mu$-fraction $\frac{2}{3}$ of all Simon inputs. Simon's
lower bound on classical algorithms now implies that
$O(T^4)=\Omega(\sqrt{2^n})$, and hence
$Q_{SE}^\mu(Simon) = \Omega(2^{\frac{n}{8}})$. 
\end{Proof}


\section{A Positive Result for the Weak Quantum Yao Principle}

In this section we show that the weak quantum Yao principle holds for 
all symmetric functions. This section is divided into three subsections, in the 
first we prove the result for threshold functions, in the second 
subsection, we extend it to symmetric functions. In the third
subsection we investigate the weak quantum Yao principle for the uniform
2-level AND-OR tree, which is monotone and non-symmetric.

\subsection{Equality up to a Constant Factor for Threshold Functions}
For every distribution $\mu$, we will exhibit a weakly $\frac{2}{3}$-exact 
quantum algorithm that computes threshold function $f$ with threshold $t$ 
in time $O(\sqrt{tN})$. This, together with Theorem \ref{thm:Q2lowerbound}
and the (known) fact that $Q_2(f)=\Theta(\sqrt{tN})$ for threshold functions
$f$ \cite{bbcmw:polynomials}, gives the desired result.

Note that given a threshold function $f:\set{0,1}^N \to \set{0,1}$ with 
threshold $t$, in order to be sure that $f(x)=1$, one will have to find 
at least $t$ 1's in the input. The crucial idea behind our algorithm is 
that if the  number of 1's in the input is large enough, then for each 
distribution $\mu$ over the inputs, we can pick a substantially smaller part
of the input such that there are between $t$ and $100t$ 1's in this 
subpart for a large $\mu$-fraction of the inputs. This idea is formally stated
in the following technical lemma.\footnote{We need the condition $i \geq 10$ in
this lemma in order to be able to approximate the hypergeometric distribution 
by a binomial distribution with sufficient accuracy.}

\begin{Lemma}
\label{lem:probability}
Let $t$ be a threshold, $\mu$ a probability distribution over
the $x \in \set{0,1}^N$,
and $i$ an integer such that $10 \leq i \leq \log N - \log t - 1$. Denote the
event $t2^i \leq |x| \leq t2^{i+1}$ by $I$, and let $x \wedge y$ denote
the bitwise AND of $x$ and $y$. There is a $y \in \{0,1\}^N$ with 
$|y|=\min \set{\frac{10N}{2^i},N}$, such that $\mathrm{Pr}_{\mu}[t \leq |x \wedge y| \leq 100t \mid I] > 0.7$.
\end{Lemma}

\begin{Proof}
Fix an $x \in \{0,1\}^N$ with $t2^i \leq |x| \leq t2^{i+1}$ and assume that
$\frac{10N}{2^i} \leq N$, for otherwise the lemma trivially holds. We claim
that if we pick a $y \in \{0,1\}^N$ with $|y|=\frac{10N}{2^i}$ uniformly at random,
then $\mathrm{Pr}[t \leq |x \wedge y| \leq 100t \mid I] > 0.7$.
To prove this claim, note that
$$
\mathrm{Pr}[|x \wedge y| = k \mid I] = 
\frac{{|x| \choose k}{N-|x| \choose |y| - k}}
{{N \choose |y|}}.
$$
This means that $|x\wedge y|$ is hypergeometrically
distributed, with expected value $E(|x\wedge y|) = \frac{|x||y|}{N}$.
Note that in this case $10t \leq E(|x\wedge y|) \leq 20t$. 
By Markov's inequality, it then follows directly that
$\mathrm{Pr}[|x \wedge y| > 100t \mid I] \leq 0.2$.

We can approximate the above distribution with a binomial distribution since
the number of draws is small compared to the size of the sample space, 
see e.g.~\cite{nicholson:normal}, and we shall henceforth
treat $|x \wedge y|$ as if it were binomially distributed, with
success probability $\theta = \frac{|x|}{N}$ and number of
draws $n=|y|$. To bound $\mathrm{Pr}[|x \wedge y| < t \mid I]$, we use 
the Chernoff bound as explained in \cite[pp.67-73]{motwani:ra}:
$$
\mathrm{Pr}[|x \wedge y| < (1-\delta)E(|x\wedge y|) \ \mid \ I] < 
e^{\frac{-\delta^2E(|x\wedge y|)}{2}}.
$$
Choosing $\delta = \frac{9}{10}$, we obtain $\mathrm{Pr}[
|x \wedge y| < t \mid I] < e^{-\frac{810t}{200}} < 0.1$.
Combining the previous two inequalities, it then follows that
$\mathrm{Pr}[t \leq |x \wedge y| \leq 100t \mid I] > 0.7$. This proves
the above claim.

Now imagine a matrix whose 
rows are indexed by the $x$ satisfying $t2^i \leq |x| \leq t2^{i+1}$ 
and whose columns are indexed by the $M = {N \choose |y|}$ different
$y$ of weight $|y| = \frac{10N}{2^i}$. We give the $(x,y)$ entry 
of this matrix value $\mu(x|I)$ if $t \leq |x\wedge y| \leq 100t$ and value 0
otherwise. By the above claim, each $x$ row will contain at least
70\% non zero entries, so the sum of the entries of each $x$ row is at
least $0.7M\mu(x|I)$. Hence, the sum of all entries in the matrix is equal
to $\sum_x 0.7M\mu(x|I)=0.7M$. But then there must be a column with
$\mu$-weight at least 0.7. The $y$ corresponding to this column is the
$y$ we are looking for in this lemma.
\end{Proof}

We will use the fact stated in the previous lemma to successively search
for $t$ 1's in exponentially smaller parts of the inputs, assuming the
presence of increasingly more 1's in the original input. The following
lemma states that this searching can be done efficiently:

\begin{Lemma}
\label{lem:count}
There exists a quantum algorithm that can find all the 1's in an input $x$ 
of size $N$ with probability 1, using at most $\frac{\pi}{2}\sqrt{kN}$ queries, 
if $k$ is a known upper bound on the number of 1's in $x$.
\end{Lemma}

\begin{Proof}
Consider Algorithm \ref{alg:ronald}. It is easily proven that this 
algorithm indeed finds all 1's, as follows. Assume an upper bound $k \geq |x|$ on the
number of 1's in $x$. If the exact version of Grover's 
algorithm finds an index of a 1 bit, then we set
this index to 0 in the search space. Because $k$ is an upper bound
on the number of 1's in $x$, we can lower $k$ each time we find
a 1, without $k$ ever becoming less than the actual number of 1's in $x$.
If it does not find a 1, then we know that our upper bound was too
high and again we can safely lower it by 1. Using these facts, it is 
easily proven by induction on $k$ that the algorithm indeed works as claimed. 

\begin{algorithm}
\caption{}
\label{alg:ronald}
\begin{algorithmic}
\FOR{$i=k$ down to 1} 
\STATE Apply Grover's exact search algorithm assuming
\STATE \hspace{10pt} there are $i$ solutions.
\IF{A solution has been found}
\STATE mark its index as a zero in the search space
\ENDIF
\ENDFOR
\STATE output the positions of all solutions found
\end{algorithmic}
\end{algorithm}

\noindent The number of queries made by this algorithm is at most:

$$\sum_{i=1}^k \frac{\pi}{4} \sqrt{\frac{N}{i}}
\leq \frac{\pi}{4}\sqrt{N} \int_{0}^{k} \frac{\mathrm{d}i}{\sqrt{i}}
= \frac{\pi}{2}\sqrt{kN}.$$
\end{Proof}

We are now ready to prove an upper bound on $Q_{WE}^\mu(f)$ for 
threshold functions.

\begin{Lemma}
\label{lem:threshold}
For threshold function $f$ with threshold $t$, and for every distribution $\mu$, 
we have $Q_{WE}^{\mu}(f) = O(\sqrt{tN})$.
\end{Lemma}

\begin{Proof}
Fix a distribution $\mu$. Invoking Lemmas \ref{lem:probability} and \ref{lem:count},
our algorithm is as follows.
First we count the number of 1's in the input
using Algorithm \ref{alg:ronald}, assuming an upper bound of $2^{10}t$ 1's. If
after that we haven't found at least $t$ 1's yet, then we successively assume
that there are between $t2^i$ and $t2^{i+1}$ 1's in the input, with $i$
going up from 10 to $\log N - \log t - 1$. For each of these assumptions,
we search a smaller part of the input. If we have reached the $i$ for
which $t2^i \leq |x| \leq t2^{i+1}$, then Lemma \ref{lem:probability} guarantees
us that for a large $\mu$-fraction of the inputs we can find a small subpart
containing between $t$ and $100t$ 1's. We then count the number of 1's in this subpart
using Algorithm \ref{alg:ronald}. Algorithm \ref{alg:thresh} is the actual
algorithm we will use.

\begin{algorithm}
\caption{}
\label{alg:thresh}
\begin{algorithmic}
\STATE Count the number of 1's in the input using Algorithm \ref{alg:ronald}, assuming an 
upper bound of $2^{10}t$ 1's
\IF{at least $t$ 1's are found}
\STATE output 1
\ENDIF
\FOR{$i=10$ to $\log N - \log t - 1$}
\STATE Let $y^{(i)} \in \{0,1\}^N$ be a string of weight $\min \set{N,\frac{10N}{2^i}}$
satisfying Lemma \ref{lem:probability}
\STATE Using Algorithm \ref{alg:ronald}, count the number of solutions in the subpart   
\STATE \hspace{10pt} of the input induced by $y^{(i)}$, assuming an upper bound of $100t$ 1's.
\IF{at least $t$ 1's are found}
\STATE output 1
\ENDIF
\ENDFOR
\STATE output 0
\end{algorithmic}
\end{algorithm}

\noindent This algorithm will be correct on all inputs $x$ with $|x| < t$ and
will produce a correct answer on at least a $\mu$-fraction 0.7 of
all inputs $x$ with $|x| \geq t$ as guaranteed by Lemma \ref{lem:probability}.
Hence it will produce a correct answer on a $\mu$-fraction of at least:
$$\mu(\{x \mid |x| < t\}) + 0.7 (1 - \mu(\{x \mid |x| < t \}) \geq 0.7.$$
Furthermore, its query complexity is equal to:
$$O(\sqrt{tN})+\sum_{i=10}^{\log N - \log t -1} O\left(\sqrt{\frac{tN}{2^i}}\right)
= O(\sqrt{tN}),$$
where the first term corresponds to the cost of searching the entire
space once with a small upper bound, and the summation corresponds to
searching consecutively smaller subparts $y^{(i)}$.
\end{Proof}

Recall that for threshold functions $f:\set{0,1}^N \to \set{0,1}$ with
threshold $t$, $Q_2(f)=\Theta(\sqrt{tN})$. By Theorem~\ref{thm:Q2lowerbound}
it then follows that $\max_\mu Q_{WE}^\mu(f) = \Omega(\sqrt{tN})$. In
combination with Lemma \ref{lem:threshold}, this yields:

\begin{Lemma}
\label{lem:eqthreshold}
For all threshold functions $f:\{0,1\}^N \to \{0,1\}$ with
threshold $t$,
$$
Q_2(f) = \Theta\left(\max_\mu Q_{WE}^\mu(f)\right) = \Theta\left(\sqrt{tN}\right).
$$
\end{Lemma}

\subsection{Equality up to a Constant Factor for Symmetric Functions.}

With the result about threshold functions in mind, we can easily prove
that the quantum Yao principle holds for all symmetric functions as well.

\begin{Theorem}
\label{thm:symmetric}
For all symmetric functions $f:\{0,1\}^N \to \{0,1\}$
$$
Q_2(f) = \Theta\left(\max_\mu Q_{WE}^\mu(f)\right) = \Theta\left(\sqrt{N(N-\Gamma(f))}\right).
$$
\end{Theorem}

\begin{Proof}
From \cite{bbcmw:polynomials} we know that
$Q_2(f)=\Theta(\sqrt{N(N-\Gamma(f))})$.
Also, Theorem \ref{thm:Q2lowerbound} tells us that
$Q_2(f) \leq \max_\mu Q_{WE}^\mu(f)$. It remains to show that
for every distribution $\mu$, $Q_{WE}^\mu(f) = O(\sqrt{N(N-\Gamma(f))})$.

Fix a probability distribution $\mu$ over the set of all inputs.
Note that $\Gamma(f)$ measures the length of the interval
around Hamming weight $\frac{N}{2}$ where $f$ is constant, so in order to
compute $f(x)$, it suffices to know $|x|$ exactly if $|x| \in
[0,\frac{N-\Gamma(f)}{2}) = I_1$ or $|x| \in (\frac{N+\Gamma(f)-2}{2},N]
= I_3$, or to know that $|x| \in [\frac{N-\Gamma(f)}{2},
\frac{N+\Gamma(f)-2}{2}] = I_2$.

We can use the threshold algorithm of the previous section to determine
whether $x\in I_1$ (with $\mu$-error probability reduced to $1/6$).
We can use another threshold algorithm to determine whether $x\in I_3$
(with the role of 0's and 1's reversed, and also with error $\leq 1/6$).
Both threshold algorithms take $O(\sqrt{N(N-\Gamma(f))})$ queries.
Now for at least $2/3$ of the inputs $x$, weighed according to $\mu$,
\emph{both} of these threshold algorithms will give the correct answer.
For all such $x$ we can determine $f(x)$ with certainty:
if we know $|x| \in I_2$ then we are done,
because $f$ is constant in this interval. If $|x|\in I_1$ or $|x|\in I_3$
then we use Algorithm \ref{alg:ronald} to count $|x|$, using
$O(\sqrt{N(N-\Gamma(f))})$ queries.
Thus we have a weakly $\frac{2}{3}$-exact quantum algorithm for $f$
with respect to $\mu$, using $O(\sqrt{N(N-\Gamma(f))})$ queries in total.
\end{Proof}

\subsection{A Result for the AND-OR Tree}\label{ssecandor}
Above we proved that the weak quantum Yao principle holds
(up to a constant factor) for all \emph{symmetric} functions.
A similar result might be provable for all \emph{monotone} functions.
Recall that a Boolean function $f$ is monotone if the function value
cannot change from 1 to 0 if we change some input bits from 0 to 1.
In this section we prove a preliminary result in this direction,
namely that the known upper and lower bounds on the $Q_2(f)$-complexity
of the 2-level \emph{AND-OR tree} carry over to weakly $\frac{2}{3}$-exact 
quantum algorithms.  This monotone but non-symmetric function is the AND 
of $\sqrt{N}$ independent ORs of $\sqrt{N}$ variables each. 
In the sequel, we use AO to denote this $N$-bit AND-OR tree.

No tight characterization of $Q_2(AO)$ is known, but
Buhrman, Cleve, and Widgerson~\cite{BuhrmanCleveWigderson98} 
proved $Q_2(AO)=O(\sqrt{N}\log N)$
via a recursive application of Grover's algorithm.
Using a result about efficient error-reduction in quantum search
from~\cite{bcwz:qerror}, this upper bound can be improved to
$Q_2(AO)=O(\sqrt{N\log N})$. This nearly matches Ambainis' lower bound 
of $\Omega(\sqrt{N})$~\cite{ambainis:lowerbounds}.
Note that Ambainis' bound together with our Theorem~\ref{thm:Q2lowerbound} 
immediately gives the lower bound
$\max_\mu Q_{WE}^\mu(AO)= \Omega(\sqrt{N})$.
Below we show that also the best known \emph{upper} bound carries over
to weakly $\frac{2}{3}$-exact algorithms: 
$Q_{WE}^\mu(AO)= O(\sqrt{N\log N})$ for all $\mu$.

To prove this result, we first show that we can efficiently reduce 
the error in weakly $\frac{2}{3}$-exact quantum search algorithms,
in analogy with~\cite{bcwz:qerror}.
For every $\mu$ and $\epsilon$, we will construct
a quantum search algorithm that uses $O\left(\sqrt{N\log(1/\epsilon)}\right)$ 
queries and solves the search problem \emph{with certainty} 
for $1-\epsilon$ of all inputs, weighed by $\mu$. 
We first need the following lemma, which states that if an
input contains many 1's, then we can deterministically reduce its
size to a smaller search space which will probably still contain at least one 1.

\begin{Lemma}
\label{lem:reduce}
For all probability distributions $\mu$ on $\01^N$ and integers $c$, 
there exists a $y\in\01^N$ with $|y|=\min\{\frac{cN}{t},N\}$, 
such that $\mathrm{Pr}_\mu[|x\land y|\geq 1\mid |x|>t]\geq 1- e^{-c}$.
\end{Lemma}

\begin{Proof}
If $\frac{cN}{t}\geq N$ then obviously the lemma holds (pick $y=1^N$),
so assume $\frac{cN}{t}<N$. Fix an $x\in\01^N$ with $|x|>t$. 
If we pick a $y\in\01^N$ with $|y|=\frac{cN}{t}$ 
uniformly at random, then
\begin{eqnarray*}
\mathrm{Pr}[|x \land y|=0\mid |x|>t] & = & \frac{{N - |y| \choose |x|}}{{N \choose |x|}}
 = \frac{(N-|x|)\cdot (N-|x|-1)\cdots (N-|x|-|y|+1)}{N(N-1)\cdots(N-|y|+1)} \\
& \leq & \left( 1 - \frac{|x|}{N}\right)^{|y|} \leq e^{-|x|\cdot |y|/N}
 \leq e^{-c}.
\end{eqnarray*}
Hence $\mathrm{Pr}[|x \land y|\geq 1\mid |x|>t] \geq 1 - e^{-c}$.
By exactly the same averaging argument as in the proof of Lemma
\ref{lem:probability}, we can show that for every distribution $\mu$, there
exists a $y$ such that $\mathrm{Pr}_{\mu}[|x \land y|\geq 1\mid |x|>t]
\geq 1 - e^{-c}$.
\end{Proof}

With Lemma \ref{lem:reduce} at our disposal, we can now prove that 
we can ``cheaply'' reduce the error of weakly $\frac{2}{3}$-exact 
quantum search algorithms to small $\epsilon$.

\begin{Lemma}
\label{lem:weaksearch}
For every $\epsilon>0$ and every probability distribution $\mu$ over $\01^N$, there
exists a weakly $(1-\epsilon)$-exact quantum search algorithm with respect
to $\mu$ that uses $O\left(\sqrt{N\log(1/\epsilon)}\right)$ queries.
\end{Lemma}
 
\begin{Proof}
Fix an error bound $\epsilon$ and distribution $\mu$.
Our $(1-\epsilon)$-exact search algorithm is inspired by~\cite{bcwz:qerror}.
Let $t_0=\log(1/\epsilon)$ (assume for simplicity that this is an integer).
First we run the exact version of Grover's algorithm 
on the input $x$ assuming that $|x|=1$, then we run it again assuming that $|x|=2$, 
and so on until $|x|=t_0$.
This takes
$$
\sum_{i=1}^{t_0}\frac{\pi}{4}\sqrt{\frac{N}{i}}=
O(\sqrt{Nt_0})=O\left(\sqrt{N\log(1/\epsilon)}\right)
$$
queries, and finds a 1 with certainty whenever $1\leq |x|\leq t_0$.

It remains to find a 1 for ``most'' of the inputs $x$ that have $|x|>t_0$.
Let $\mu_1$ be the probability distribution $\mu$ restricted to the $x$
with $|x|>t$. By Lemma~\ref{lem:reduce}, we know there exists a $y\in\01^N$
with $|y|=O(N/t_0)$ such that 
$\mathrm{Pr}_{\mu_1}[|x\land y|\geq 1]=\mathrm{Pr}_\mu[|x\land y|\geq 1\mid |x|>t_0]
\geq\frac{5}{6}$. Now we use a $\frac{2}{3}$-exact quantum search algorithm
with respect to $\mu_1$ to search the subpart of $x$ indicated by $y$. 
This subpart has size $O(N/t_0)$, and Lemma~\ref{lem:threshold} guarantees
us that there is $\frac{2}{3}$-exact algorithm with $O(\sqrt{N/t_0})$ queries.
Thus we find a 1 with certainty for a $\mu_1$-fraction (and hence also
$\mu$-fraction) of at least $\frac{5}{6}-\frac{1}{3}=\frac{1}{2}$ 
of the inputs with $|x|>t_0$.
Now we repeat this idea to ``catch'' $\frac{1}{2}$ of the remaining inputs.
Let $\mu_2$ be $\mu_1$ restricted to the inputs with $|x|>t_0$ where the 
previous algorithm did not find a 1 with certainty. Using another
$\frac{2}{3}$-exact algorithm (this time with respect to $\mu_2$)
for another $y$, we can catch $\frac{1}{2}$ of the remaining inputs.
We repeat this $t_0$ times and eventually catch
$$
1-\left(\frac{1}{2}\right)^{t_0}= 1-\epsilon
$$ 
of the inputs (weighed according to $\mu$) in this way.
If we still have not found a 1 after all this, we stop and output 
`no solutions', which ensures that our algorithm is always correct on
the all-0 input.
Note that the second part of the algorithm uses 
$$
t_0\cdot O(\sqrt{N/t_0})=O\left(\sqrt{N\log(1/\epsilon)}\right)
$$
queries, so our overall query complexity is 
$O\left(\sqrt{N\log(1/\epsilon)}\right)$, as promised.
\end{Proof}

Using Lemma \ref{lem:weaksearch} we now show that the best 
known upper bound for $Q_2(AO)$ also holds for 
weakly $\frac{2}{3}$-exact quantum algorithms.

\begin{Theorem}
For every distribution $\mu$ on $\{0,1\}^N$ we have $Q_{WE}^\mu(AO)=O(\sqrt{N\log N})$. 
\end{Theorem}
 
\begin{Proof}
Fix some distribution $\mu$. We will sketch a $\frac{2}{3}$-exact
quantum algorithm for AO with respect to $\mu$, along the lines
of the recursive-Grover of~\cite{BuhrmanCleveWigderson98}.
For each of the $1\leq i\leq\sqrt{N}$ OR functions at the ``bottom''
of the tree, let $\mu_i:\{0,1\}^{\sqrt{N}}\rightarrow[0,1]$ 
be the distribution over its $\sqrt{N}$ input bits induced by $\mu$,
i.e., $\mu_i(y)$ is the sum of $\mu(x)$ over all $x\in\{0,1\}^N$ where 
the $i$th block of $\sqrt{N}$ variables takes value $y$.
Let $A_i$ be a weakly $\left(1-\frac{1}{6\sqrt{N}}\right)$-exact 
quantum algorithm with respect to $\mu_i$ for the $i$th OR. 
By Lemma \ref{lem:weaksearch}, each $A_i$ takes 
$O\left(\sqrt{\sqrt{N}\log N}\right)$ queries.
Note that now for $\frac{5}{6}$ of the inputs, weighed according to $\mu$,
\emph{all} $A_i$ deliver the correct answer with certainty.
By standard techniques (copying the answer and reversing the computation 
afterwards~\cite{bernstein&vazirani:qcomplexity,cdnt:ip}) 
we can ``clean up'' these computations, setting the workspace
back to the initial state and just retaining the answer bit.

We now want to run a $\frac{5}{6}$-exact quantum algorithm for AND
on top of these $\sqrt{N}$ subtrees to compute the AND-OR tree.
Let $\mu':\{0,1\}^{\sqrt{N}}\rightarrow[0,1]$ be the induced input 
distribution for the top-AND, i.e., $\mu'(y)$ is the sum of $\mu(x)$ 
over all $x\in\{0,1\}^N$ where the $i$th OR takes the value $y_i$,
for all $1\leq i\leq \sqrt{N}$.
Let $A$ be a weakly $\frac{5}{6}$-exact quantum algorithm for 
the $\sqrt{N}$-variable AND with respect to $\mu'$.
By Lemma~\ref{lem:threshold} such an algorithm makes $O(N^{1/4})$ queries.
If we replace, in $A$, a query to the $i$th bit by a call to $A_i$,
then we obtain an $O(\sqrt{N\log N})$-query algorithm
that is correct with certainty on a $\mu$-fraction at least
$\frac{5}{6}-\frac{1}{6}=\frac{2}{3}$ of all inputs.
\end{Proof}

\section{Summary and Open Problems}

In this paper we investigated to what extent quantum versions
of the classical Yao principle hold. We formulated a strong and a weak
version of the quantum Yao principle, showed that both hold in one direction,
falsified the other direction for the strong version, and
proved the weak version for the query complexity of all symmetric functions.

The main question left open by this research is the general validity of the
weak quantum  Yao principle. 
On the one hand, we may be able to find a counterexample
to the weak principle as well, perhaps based on the query
complexity of the \emph{order-finding problem}.
Shor showed that the order-finding problem can be solved by a bounded-error quantum 
algorithm using $O(\log N)$ queries~\cite{shor:factoring}.
Using Cleve's $\Omega(N^{1/3}/\log N)$ lower bound on classical algorithms
for order-finding~\cite{cleve:orderfinding}, we can exhibit a $\mu$
such that any strongly $\frac{2}{3}$-exact
quantum algorithm for $f$ with respect to $\mu$ requires $N^{\Omega(1)}$ queries
(in the same way as Theorem~\ref{thm:Q2lowerbound}).
This gives another counterexample to the strong quantum Yao principle.
The same problem may even provide a counterexample to the \emph{weak}
quantum Yao principle, as it seems hard to construct even weakly
$\frac{2}{3}$-exact quantum algorithms for this problem.

On the other hand, we may try to extend the class of functions
for which we know the weak quantum Yao principle \emph{does} hold.
A good starting point here might be the class of all \emph{monotone}
functions. We discussed one such function, the 2-level AND-OR tree,
in Section~\ref{ssecandor}.
Unfortunately, at the time of writing no general characterization
of the $Q_2(f)$-complexity of all monotone functions is known,
in contrast to the case of symmetric functions. 
Also, in this direction it might be a fruitful idea to further 
explore the rapidly growing field of \emph{quantum game theory} 
(see for example \cite{meyer:games}) and the possible connections 
between that area and our work.

\section*{Acknowledgments}
We thank Harry Buhrman for initiating this research, for coming up with
the counterexample of Theorem~\ref{thm:exponentialGap}, and for useful
comments on a preliminary version of this paper. We thank him
and Peter H\o yer for their contributions to an initial proof of the
weak quantum Yao principle for the OR function, which forms the basis
for the current proof of Theorem~\ref{thm:symmetric}. We also thank Leen
Torenvliet for useful discussions.


\end{document}